\documentclass[preprint,aps,nofootinbib]{revtex4}
\usepackage{graphicx}% Include figure files
\usepackage{dcolumn}% Align table columns on decimal point
\usepackage{bm}% bold math
\usepackage{epsf}
\usepackage{float}
\def\bq{{\bf q}}

\newcommand{\edc}{\end{document}}
\newcommand{\bb} {\begin{thebibliography}}
\newcommand{\eb}{\end{thebibliography}}
\newcommand{\bi}[1]{\bibitem{#1}}
\newcommand{\bc}{\begin{center}}
\newcommand{\ec}{\end{center}}

\newcommand{\be}{\begin{equation}\small}
\newcommand{\ee}{\end{equation}\normalsize}
\newcommand{\bea}{\begin{eqnarray}}
\newcommand{\eea}{\end{eqnarray}}
\newcommand{\ba}{\begin{array}{l}   }
\newcommand{\lab}[1]{\label{#1}}
\newcommand{\ea}{\end{array}}

\newcommand{\dsfrac}{\displaystyle\frac}
\newcommand{\ds} {\displaystyle}
\newcommand{\summa}{\ds\sum}
\newcommand{\dssum}{\summa}
\newcommand{\re}[1]{(\ref{#1})}

\newcommand{\ci}{\cite}

\newcommand{\dsint}{\ds\int}

\def\bfr{{\bf r}}
\def\bfk{{\bf k}}
\def\bfq{{\bf q}}

\newcommand{{\vergul}}{  ,}

\newcommand{\veps}{\varepsilon }

\begin{document}
%\twocolumn
%\sloppy
\draft
%\doublespace
\title{Critical temperature of noninteracting bosonic gases in cubic optical lattices at arbitrary integer fillings}
\author{Abdulla Rakhimov$^{a,b}$},\email{rakhimovabd@yandex.ru}
\author{ Iman N. Askerzade $^{b,c}$}
\address{
$^a$ Institute of Nuclear Physics, Tashkent 100214, Uzbekistan\\
$^b$ Computer Engineering  Department  of Ankara University, Ankara, 06100, Turkey\\
$^c$ Institute of Physics Azerbaijan National Academy of Sciences, AZ1143, Baku, Azerbaijan\\
}
%\ead{rakhimovabd@yandex.ru}

%%%%%%%%%%%%%%%%%%%%%%%%%%%%%%%%%

%%%%%%%%%%%%%%%%%%%%%%%%%%%%%%%%%
\begin{abstract}

We have shown that the critical temperature of a Bose-Einstein condensate to a normal phase transition
of noninteracting bosons in cubic optical lattices has a linear dependence on the filling factor,
especially at large densities. The condensed fraction exhibits a linear power law dependence
 on temperature in contrast to the case of ideal homogeneous Bose gases.

\end{abstract}


\maketitle

%%%%%%%%%%%%%%%%%%%%%%%%%%%%%%%%%%%%
\section{Introduction}
%%%%%%%%%%%%%%%%%%%%%%%%%%%%%%%%%%%%
 Ultracold bosonic atoms in optical lattices have sparked investigations of strongly
 correlated many-body quantum phases with ultracold atoms
 \ci{lewenstein} that are now at the forefront of current research.
 They may be used as quantum emulations of a more complex condensed matter system.
Experimentally they are created  by superimposing two counter-propagating laser beams
of the same wavelength and frequency that act as an periodic potential. In the simpliest
case, when the depth is constant and isotropic the potential can be represented as follows:
 \be
 V_{L}(\bfr)=V_{0}\sum_{\alpha=1}^{d}\sin^{2}(k_{0}^{\alpha}r_{\alpha}),\lab{1.1}
 \ee
where the wave vector $\bfk_{0}\equiv\{k_{0}^{\alpha}\}$ is related to the laser wavelength
$\lambda_{\alpha}$ as $k_{0}^{\alpha}=2\pi/\lambda_{\alpha}$, and $d$ is the space dimension
of a cubic lattice, $d=1,2,3$.

It is well known that, an ideal homogenious  Bose (IHB) gas  of noninteracting atoms
consists of free atoms with the  plane wave
$\exp(i\bfk \cdot \bfr)$, and with the  energy dispersion relation $\epsilon_{k}=\bfk^{2}/2m$.
The creation of an optical lattice may be considered as a procedure of loading
preliminarily magnetically trapped ultracold Bose atoms into a well tuned
laser field, whose influence on the atoms, being in fact  the  Stark effect, is simulated via
a periodic potential \re{1.1}. Now the dispersion is no longer quadratic with
the momentum, but develops gaps at specific locations determined by the lattice structure.
This energy can be specified by a band index and a quasimomentum, taking on
values within the first Brillouin zone only.  As to the wave function it can be written as a Bloch
function
$\chi_{nk}(\bfr)=\sum_i \exp(i\bfk \cdot \bfr)\omega_n(\bfr-\bfr_i)$
 in the Wannier representation.
In the limit $V_0\gg E_R$, where $E_R$ is the recoil energy, each well of the periodic potential supports a number of vibrational
levels, separated by an energy $\omega_0\gg E_R$. At low temperatures, atoms are restricted to the vibrational level at each site. Their kinetic energy is then frozen,except
for the small tunneling amplitude to neighboring sites. The associated single particle
eigenstates in the lowest band are Bloch waves with quasimomentum
${\bf q}$ and energy $\epsilon_0(q)=3/2\omega_0-2J[\cos(q_x a)+\cos(q_y a)+\cos(q_z a)]+\ldots$,
where $\omega_0$ is the energy of local oscillations in the well \ci{zwerger}.
This is one of the main differences between IHB gas and noninteracting Bose gas
in optical lattices, which is no longer homogenious either.
The bandwidth parameter $J>0$ is the gain in the  kinetic energy due to the nearist
neighbor tunneling, which can be approximated for $d=3$ as

\be
  J\approx\dsfrac{4}{\sqrt{\pi}}
  E_{R}\exp
  \left
  [-2\sqrt{\dsfrac{V_{0}}{E_{R}}}
     \right
    ] \left[
{\dsfrac{V_{0}}{E_{R}}}
\right ]^{3/4}
    \lab{2.1}
     \ee
where  $E_{R}= k_{0}^{2} / {2m}   $ , $k_0=\mid {\bf k}_0 \mid $ is the laser
 wave vector modulus, and  $k_0=\pi/a$, $a$ is the lattice spacing.

By the assumption that   only the lowest band is taken into account, an optical
lattice without a harmonic  trap can be described by the Bose-Hubbard model \ci{stoofbook},
\be
 \hat{H}=-J\sum_{\langle i,j\rangle}(\hat{b}_{i}^{+}\hat{b}_{j}+hc)+\frac{U}{2}\sum_{j}^{N_s}\hat{n}_{j}
  (\hat{n}_{j}-1),\lab{1.2}
   \ee
where $\hat{b}_{i}^{+}$   and $\hat{b}_{i}$ are the bosonic creation and annihilation operators on the site $ i$; the sum over $\langle i,j\rangle$ includes only
pairs of nearest neighbors; $J$ is the hopping amplitude, which is responsible for the tunneling of an atom from one
site to another neighboring site; $U$ is the on site repulsion energy, $\hat{n}_{j}$ is the number operator and $N_s$ the number of sites.
 Depending on the ratio $\kappa=U/J$ , the filling factor $\nu$ and the temperature $T$,  the system may be in superfluid (SF),
 Mott insulator (MI) or in normal (N) phases. Note that, strictly speaking MI phase may be reached
 only for $T=0$ and  commensurate, i.e. integer filling factors \ci{Yukalovobsor}, $\nu=N/N_{s}$, where $N$ is the total
 number of atoms. The filling factor is related to
  the average atomic density, $\rho=N/V$ as $\nu=\rho a^d$, where $V$ - volume of the system occupied with the atoms.

 The critical temperature $T_{c}$ of the phase transition is determined by the filling factor, the interatomic coupling  strength and  the parameters of external harmonic trap. In general,  $T_{c}$ may
  be considered as a sum of following three terms \ci{bailie}
  \be
  T_{c}=T_{c}^{0}+\Delta T_{c}^{HT}+\Delta T_{c}^{U},\lab{2.2 oi}
   \ee
   where $\Delta T_{c}^{HT}$ and $\Delta T_{c}^{U}$ are the shifts due to the trap and interatomic
   interactions,  respectively, and $T_{c}^{0}$  is the critical temperature for the ideal case corresponding to
   the Hamiltonian with $U=0$ in Eq. \re{1.2}. Presently $T_{c}$ has been estimated in quantum Monte Carlo (QMC) simulations for
   $d=2$ \ci{mahmud} and for $d=3$ \ci{svistun}  as well as experimentally  \ci{trotzky}.
   These calculations, including the ones, based on the mean field  approximation \ci{ourknr2,ouryukalov},
   have
   shown the down shift of $T_{c}$ $[$ i.e. $\Delta T_{c}=(T_c-T_{c}^{0})/T_{c}^{0}
   <0$ $]$ at large $U/J$.
   Note that, the majority of calculations including $\Delta T_{c}^{HT}$ as well as
   $\Delta T_{c}^{U}$ are performed for small $\nu, (\nu\leq 3)$ \ci{bailie}, so the
   dependence of even $T_{c}^{0}(\nu)$ on $\nu>3$ is poorly known. Thus evaluation of $T_{c}^{0}(\nu)$
     for various filling factors may be a subject of separate studies .

   In the present work we will calculate $T_{c}^{0}(\nu)$ for $d=1,2,3$ for arbitrary integer filling factors and make
   an interpolation of the  $T_{c}^{0}(\nu)$ function. Since an ideal gas with the quadratic  spectrum at small momentum, i.e. $\epsilon(k)=ck^2+O(k^3)$  can not exhibit
   superfluidity \ci{landau}, our  discussions will concern  the phase transition from the Bose Einstein condensate (BEC)
   phase into a normal phase. Although, strictly speaking,
    an ideal gas is an unstable system with anomalously
strong particle fluctuations \ci{yuklpl}, the present work will give an opportunity for
    estimations in future  experiments  and QMC calculations for large values of  $\nu$ and may serve as
   a check point in theoretical studies in the limit $(U/J) \rightarrow 0 $.  In Secs. II and III we  study the critical temperature
   $T_{c}^{0}(\nu)$ and the condensed fraction $n_0$ respectively. In Sec. III we
   compare some scaling properties of thermodynamic quantities of noninteracting Bose atoms in optical lattices   with that of IHB gas also. The results will be briefly summarized in Sec. IV.
   Below we use $\hbar=1$, $k_B=1$.

\section{The critical temperature}

   The density of a homogeneous  ideal atomic Bose gas with the chemical potential $\mu$ is given by
   \be
    \rho=\frac{N}{V}=\int \dsfrac{d^{d}k}{(2\pi)^{d}(e^{(\epsilon_{k}-\mu)/T}-1)},
    \lab{3.1}
     \ee
where $\epsilon_{k}=\bfk^{2}/2m$. The phase transition BEC$\to$N occurs when
$\mu=0$, so that the critical temperature ${\tilde T}_{c}^{0}$ is determined by the  following equation with a given density $\rho_{c}$:
\be
 \rho_{c}=\int \dsfrac{d^{d}k}{(2\pi)^{d}(e^{\epsilon_{k}/{\tilde T}_{c}^{0} }-1)},\lab{3.2}
  \ee
   which may be formally  evaluated analytically to give
  \be
  {\tilde T}_{c}^{0}=\dsfrac{2\pi}{m}\left[\dsfrac{\rho_{c}}{g_{d/2}(1)}    \right]^{2/d}
  \lab{tc23}
  \ee
  where $g_{p}(z)=1/\Gamma(p)\int_{0}^{\infty} x^{(p-1)}dx/(\exp (x)z^{-1}-1)  $ is a Bose function.
   However, when $d\leq2$  the momentum integration in
  \re{3.2} has an  infrared divergency at small $k$, and can not be evaluated. Physically this means
  that for $d\leq2$ the chemical potential  can not reach zero; in other words, there
   would be no phase transition from BEC into a normal phase at finite temperatures \ci{dalfavo}.
   Nevertheless, in one and two dimensions  the systems may exhibit superfluidity or Berezinski-
   Kosterlitz-Thouless (BKT) phase transition, which may take place at a critical temperature ${\tilde T}_{c}^{0}$.

   Now we turn to a case of periodic system of ideal, i.e. , noninteracting  Bose gases in optical lattices, which are described
   by the Hamiltonian \re{1.2} with $U=0$ and $ J\neq0$.
   This system  is discrete with a lattice spacing $a_{i}=
   \lambda_{i}/2$ and contains a finite number of atoms, $N<\infty$ . Since the optical potential is periodic, $V_{L}(\bfr+{\bf a}_{i})=V_{L}(\bfr)$,
   one may employ the Bloch functions, $\varphi_{nk}(\bfr)=e^{i\bfk\bfr}f_{nk}(\bfr)$ labeled by
   the band index $n$ where the  quasimomentum $\bfk$ pertains to the Brillouin zone $\mathcal{B}=\{\bfk:
   -\pi/a_{i}\leq k_{\alpha}\leq\pi/a_{i}\}$. The number of $k$-points  in the Brillouin zone equals to the number
   of sites: $N_{s}\equiv V a^{-d}=\dssum_{k\in\mathcal{B}} (1)$. Thus, assuming that, the thermodynamic limit
   holds for a such system for arbitrary $\nu$, one may rewrite Eq. (\ref{3.2}) as follows: \ci{ourknr1, danshita}
   \be
    \nu=\frac{1}{N_{s}^{d}}\dssum_{q_{1}=1}^{N_{s}-1}\dssum_{q_{2}=1}^{N_{s}-1}\ldots
     \dssum_{q_{d}=1}^{N_{s}-1}\dsfrac{1}{e^{\varepsilon(\bq)/T_{c}^{0}}-1},\lab{4.1}
      \ee
   with the bare dispersion
   \be
    \varepsilon(\bfq)=2J\left(d-\sum_{\alpha=1}^{d}\cos(2\pi q_{\alpha}/N_{s})\right)\lab{4.2}
     \ee
 Note that the $\bq=0$ mode, i.e., Goldstone mode  is omitted from the sum due to the orthogonality
 between the condensed and noncondensed modes due to the prescription given by Danshita and Naidon
  \ci{danshita}.
 For $d=3$,  the function $q^{2}/[\exp(\varepsilon(\bq)/T_{c}^{0})-1]$ is regular
  at $q=0$, and in the thermodynamic limit ($N_{s}\to\infty$) the summation in \re{4.1} may be replaced by the integration
  as follows \ci{ourknr2,ouryukalov}
  \be
   \nu\simeq\dsint_{0}^{1}dq_{1} dq_{2}dq_{3}\dsfrac{1}{e^{\varepsilon'(\bq)/T_{c}^{0}}-1}\lab{4.3},
    \ee
    with  $\varepsilon'(\bq)=2J\dssum_{\alpha=1}^{3}(1-\cos\pi q_{\alpha})$.

    In numerical calculations it is convenient to rewrite (\ref{4.1}) in following dimensionless form
    \be
    \nu=\frac{1}{N_{s}}\sum_{q}\frac{1}{e^{2\tilde{\varepsilon}({\bq})/\nu t_{c}^{0}(\nu)d}-1}
    ,\lab{4.4}
      \ee
where we presented $T_{c}^{0}$ and $\varepsilon(\bfq)$ as
\be
 T_{c}^{0}=\nu d J t_{c}^{0}(\nu), \quad
  \varepsilon(\bfq)=2J\tilde{\varepsilon}(\bfq)\lab{4.5}.
   \ee
and introduced the notation
\be
     \frac{1}{N_{s}}\sum_{q}f(\bq)\equiv\frac{1}{N_{s}^{d}}\sum_{q_{1}=1}^{N_{s}-1}\sum_{q_{2}=1}^{N_{s}-1}\ldots
     \sum_{q_{d}=1}^{N_{s}-1}f(q_1,q_2,..q_d)
\ee
Now for  given parameters, $N_{s}\geq60$, and $d$, one may perform numerical summation (or integration) in (\ref{4.4})
 and evaluate $t_{c}^{0}(\nu)$ for any filling factor $\nu$. In Fig.1(a)
 we present  $t_{c}^{0}(\nu)$ as an exact solution to the equation (\ref{4.4}).
 It is seen that $t_{c}^{0}(\nu)$ reaches its asymptotic value (shown by the dashed line) even at $\nu \sim 10$ where it  becomes constant.

To understand such behavior of the function $t_{c}^{0}(\nu)$ in detail    we consider the case with a large filling factor,
 $\nu \gg 1$. Approximating the exponential function  in \re{4.4} as $e^x\approx 1+x$  and solving the resulting equation  with respect to
 $t_{c}^{0}(\nu)$ one finds \footnote{Although this  simple approximation is not valid for
 the system of  homogenous
 atomic gases, it is justified for optical lattices due to the fact that
 $\tilde{\varepsilon}(\bfq)$ is bounded above i.e. $\mid \tilde{\varepsilon}(\bfq) \mid \leq 2d$.  }
\be
  \lim_{\nu\to\infty}t_{c}^{0}(\nu)=\frac{2}{d C_d}\equiv a\lab{5.1},
   \ee
   where
   \be
   C_d=\displaystyle{\frac{1}{N_{s}}\sum_{q}
     \dsfrac{1}{\tilde{\varepsilon}(\bq)}
          }
          \lab{cq}
     \ee
      is a $C$ - number, (see Table 1). Therefore, from equations \re{4.5} and \re{5.1} one may conclude that
      in the limit $\nu \gg 1$,  $T_{c}^{0}(\nu)$ is  linear in $\nu$, i.e.  $T_{c}^{0}(\nu)\approx2J\nu/C_d$.
      The linear dependence of the function $T_{c}^{0}(\nu)$ on $\nu$ was also shown by Yukalov
      \ci{Yukalovobsor} in Debye - like
      approximation, when the integral over the Brillouin
zone is replaced by the Debye sphere, whose
radius is chosen so that to retain the normalization condition
$\int _{\cal B}d \vec{k}/(2\pi)^{d}=\rho/\nu$.

%    The solutions of the Eq. (\ref{4.4}) for $d=1,2,3$ in a large scale of $\nu$ are presented in %Fig.1a.
     It is seen from Fig.1(a) that $t_{c}^{0}(\nu)$ reaches  its asymptotics given by \re{5.1}
     at rather small values of $\nu$.
     For $\nu>3$ this function may be approximated as
     \be
      t_{c}^{0}(\nu)=ae^{b/\nu}\lab{6.1},
       \ee
   and hence
   \be
    T_{c}^{0}=\nu J d   ae^{b/\nu},\lab{6.2}
     \ee
   where the fitted values of $b$ for   various $d$ are presented in Table 1.
   \begin{table}[ tp]%
\caption {The paramters of the function  $T_{c}^{0}=\nu J d   ae^{b/\nu}$. The last
column presents numerical values of $C_d$ evaluated by Eq. \re{cq}.}
\begin{tabular}{|c|c|c|c|}
  \hline
   $d$         &   a   &    b & $C_d$ \\
  \hline
  $1$         & 0.15 & 0.19 & 13.33 \\
  \hline
  $2$         & 0.86  & 0.33 & 1.16\\
  \hline
  $3$         & 1.32  & 0.37  &  0.505\\
  \hline
  \end{tabular}\lab{tab1}
\end{table}

   In Fig.1(b) the critical temperatures $T_{c}^{0}(\nu)$  (in units $J$) for noninteracting  Bose gases in optical lattices  versus filling factor $\nu$, evaluated exactly from
   Eq.s (\ref{4.4}) and (\ref{4.5}), are presented. It may be concluded  that the critical temperature
   of ideal optical lattice has a rather linear dependence on  $\nu$, which plays the role of the density,
   than IHB gas with the $(2/3)$ power law dependence given by Eq.  \re{tc23}.
   %%%%%%%%%%%%% FIGURE 1A,B %%%%%%%%%%%%%%%%%%%%%%%%%%
\begin{figure}[h]
\begin{minipage}[h]{0.49\linewidth}
\center{\includegraphics[width=1.1\linewidth]{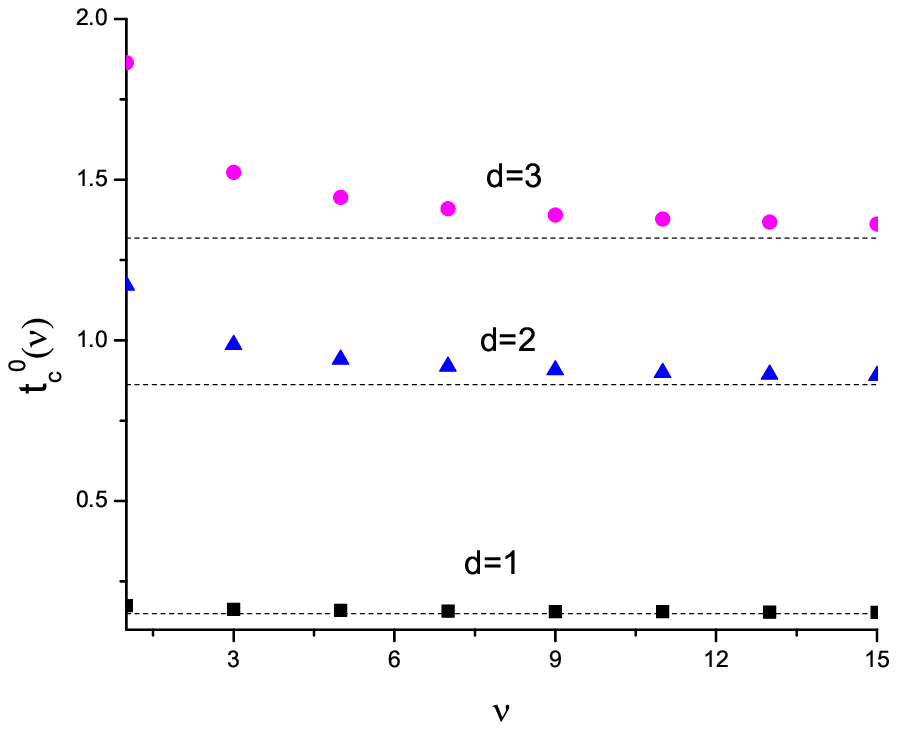} \\ a)}
\end{minipage}
\hfill
\begin{minipage}[h]{0.49\linewidth}
\center{\includegraphics[width=1.1\linewidth]{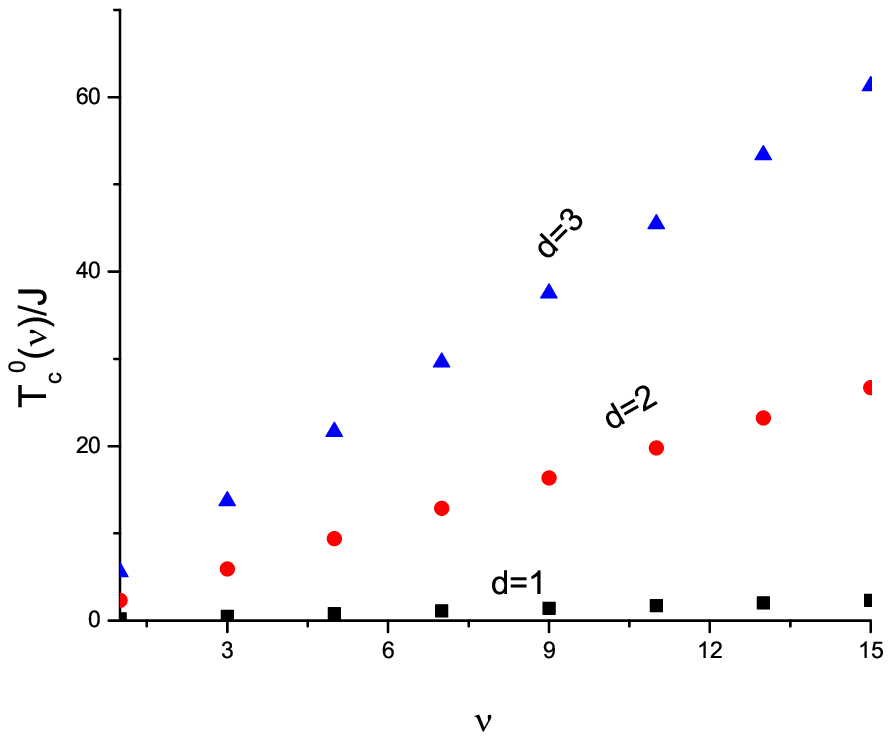} \\ b)}
\end{minipage}
\caption{(Color online) The solution of Eq. \re{4.4}
$t_{c}^{0}(\nu)$ (a)  shown as symbols, and the dimensionless critical temperature, $T_{c}^{0}/J$,
 (b) of ideal cubic optical  lattice versus the filling
 factor $\nu$  in realistic dimensions $d=1,2,3$. The dashed lines in
 panel (a) show the asymptotics of $t_{c}^{0}(\nu)$ given by \re{5.1}.  \label{fig:1}}.
\end{figure}
%%%%%%%%%%%%%%%%%%%%%%%%%%%%%%%%%%%%%%%%%%%%%%%%%%%%%%%
\section{The condensed fraction and scaling}

It is well known that for IHB gas in the  $d=3$ dimension the critical behavior of the condensed fraction
${\tilde n}_0(T)$, is given by a $(3/2)$ power law:
\be
{\tilde n}_0(T)=1-\left [ \frac{T} {\tilde{T}_{c}^{0}  }\right ]^{3/2}
\lab{n032}
\ee
As for the energy $E(T)$ and the heat capacity $C_v (T)=(\partial E(T)/\partial T)_V$ in  the condensed phase, they increase
with the temperature  as $E(T)\sim T^{5/2}$ and  $C_v (T)\sim T^{3/2}$ \ci{landau}.
The question arises, do these power laws hold for the system of ideal Bose gases in optical lattices as well?
 Below we will show that in this case, especially at large filling factors, the system preferes a linear dependence of $n_0(T)$ and $E(T)$ rather than
 $3/2$ or $5/2.$

 Actually, the corresponding condensed fraction $n_0(T)$ may be presented as
 \be
 \ba
 n_0(T)= 1- \dsfrac{1}{\nu N_s}\dssum_q \dsfrac{1}{e^{\varepsilon(\bq)/T}-1}
 \equiv\\
 \\
 \dsfrac{1}{\nu N_s}\dssum_q \dsfrac{1}{e^{\varepsilon(\bq)/T_{c}^{0}}-1}-\dsfrac{1}{\nu N_s}\dssum_q \dsfrac{1}{e^{\varepsilon(\bq)/T}-1}
  \lab{n0t}
 \ea
 \ee
 In the large $\nu \gg 1$ limit one may use the linear approximation in the exponential
 functions in \re{n0t} to obtain a relation similar to \re{n032}. In fact, introducing the reduced temperature
 $t$ as $t=T/ T_{c}^{0} $, $0\leq t\leq 1$, we have obtained
 \be
 \ba
 n_{0}(T)\mid_{\nu \gg 1}\approx\dsfrac{1}{\nu N_s}\dssum_{q} \left[\dsfrac{d\nu t_{c}^{0}(\nu)}{2\tilde\veps (\bq)}
 - \dsfrac{td\nu t_{c}^{0}(\nu)}{2\tilde\veps (\bq)} \right ]
 =\\
 \\
 1-t=1-\dsfrac{T}{ T_{c}^{0}}
 \lab{n0tlim}
 \ea
 \ee
 where we have used Eqs.  \re{4.5} and \re{5.1}. Therefore at large $\nu$
the condensed fraction  exhibits a linear power law at $T\leq T_{c}^{0}$ as $n_0\vert_{\nu\to\infty}=1-T/{ T_{c}^{0}}$
in contrast to Eq. \re{n032}.
The  linearity  of the  function $E(T)/N$, i.e. the energy per atom in the BEC phase, defined as
\footnote{Here the vacuum energy, $E(T=0)/N$, has been subtracted. }
\be
\dsfrac{E(T)}{N}=\dsfrac{1}{ \nu N_s}\dssum_q \dsfrac{\varepsilon(\bq)}{e^{\varepsilon(\bq)/T}-1}=\dsfrac{T}{\nu}+O(\frac{T^2}{\nu^2}),
\lab{et}
\ee
  at large filling factors and hence,  the condition
  \be
\ds\lim_{\nu\to\infty}C_v(T)=N_s
  \ee
 may be proven in  a similar way.

In Fig. 2  the condensed fraction $n_0(T)$ vs $T/T_{c}^{0}$, found as an exact solution to
  Eq. \re{n0t}, is presented for various values of $\nu$. To illustrate the contrast between
  the ideal gases,
satisfying  Eq. \re{n032} and the  optical lattices
we plot here the $(3/2)$ power law dependence  also (solid thick line).
It is seen that the function $n_0(T)$ exhibits a linear dependence for $\nu\geq 5$.

The energy  $E(t)/N_s$ in units $ T_{c}^{0}$ and the heat capacity $C_
v (t)/N_s$ for different values of
$\nu$ are plotted in Figs.3 and 4, respectively. It is seen that the energy becomes linear
on temperature,   and the corresponding heat capacity, $C_v/N_s$ reaches its asymptotics, i.e.  unity even at $\nu \sim 5$, $t\sim 0.4$ . On the other hand, it is well known that , e.g., heat capacity per particle of
 IHB gas,
is  a function of only the reduced temperature, and does not depend on the density, i.e.
$[{C_v/N}]_{gas}\equiv [{C_V/N}]_{gas} (t)$. But, for the case of noninteracting gas in the periodic potential, one may note from Fig.4 that, its heat capacity depends on the filling factor also, which plays the role of the density. This is one of the main results of the present work. This dependence
may be approximated as
\be
\dsfrac{C_v(t,\nu)}{N_s}=[1-\dsfrac{1}{e^{2.286\nu}}]t^{0.716/\nu}
\lab{appcv}
\ee
in the BEC regime, i.e. for $t\leq 1.$ The heat capacity per particle may be found from \re{appcv}
by using the formula $ C_v(t,\nu)/N=[C_v(t,\nu)/N_s]/\nu$.

%%%%%%%%%%%%% FIGURE 2 %%%%%%%%%%%%%%%%%%%%%%%%%%
\begin{figure}[h!]
\begin{center}
\leavevmode
\includegraphics[width=0.5\textwidth]{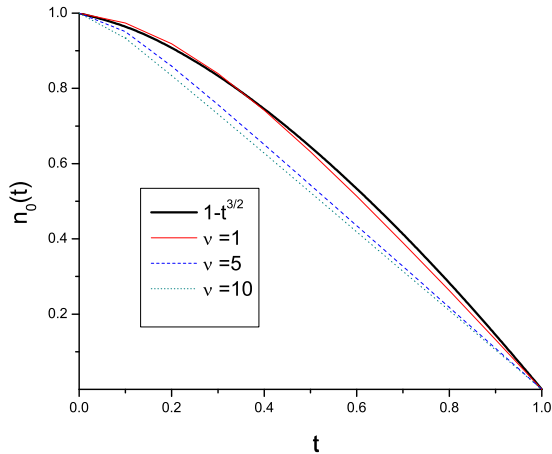}
\end{center}
\caption{(Color online)
The condensed fraction $n_0$ vs reduced temperature $t=T/T_{c}^{0}$, as an exact solution of Eq. \re{n0t}
for $d=3$ and different values of $\nu$. The thick solid line corresponds to the IHB gas.
 }
\label{fig2}
\end{figure}

%%%%%%%%%%%%% FIGURE 3 %%%%%%%%%%%%%%%%%%%%%%%%%%
\begin{figure}[h!]
\begin{center}
\leavevmode
\includegraphics[width=0.5\textwidth]{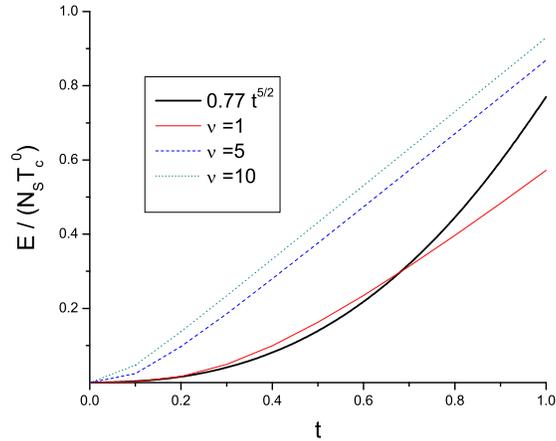}
\end{center}
\caption{(Color online)
 The total energy per site in units $T_{c}^{0}$ vs reduced temperature $t=T/T_{c}^{0}$,
for $d=3$ and different values of $\nu$.  The thick solid line corresponds to the IHB gas.
 }
\label{fig3}
\end{figure}

%%%%%%%%%%%%% FIGURE 4 %%%%%%%%%%%%%%%%%%%%%%%%%%
\begin{figure}[h!]
\begin{center}
\leavevmode
\includegraphics[width=0.5\textwidth]{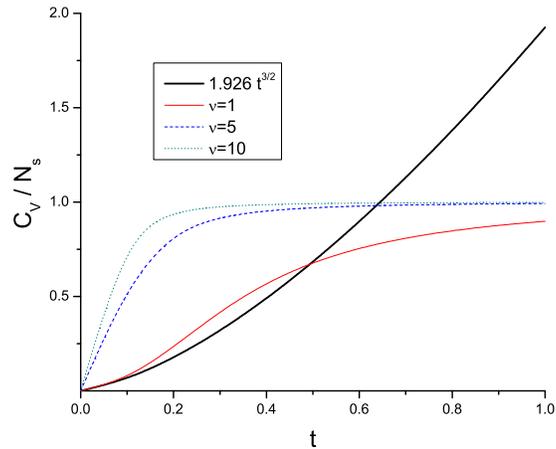}
\end{center}
\caption{(Color online)
The same as in Fig.3 but for the heat capacity, $C_v/N_s$. It seen that at
high  filling factors, $\nu\geq 10$, the  $C_v/N_s$ becomes constant at $T\geq 0.1 T_{c}^{0}$
 }
\label{fig4}
\end{figure}

%%%%%%%%%%%%% FIGURE 5 %%%%%%%%%%%%%%%%%%%%%%%%%%
\begin{figure}[h!]
\begin{center}
\leavevmode
\includegraphics[width=0.5\textwidth]{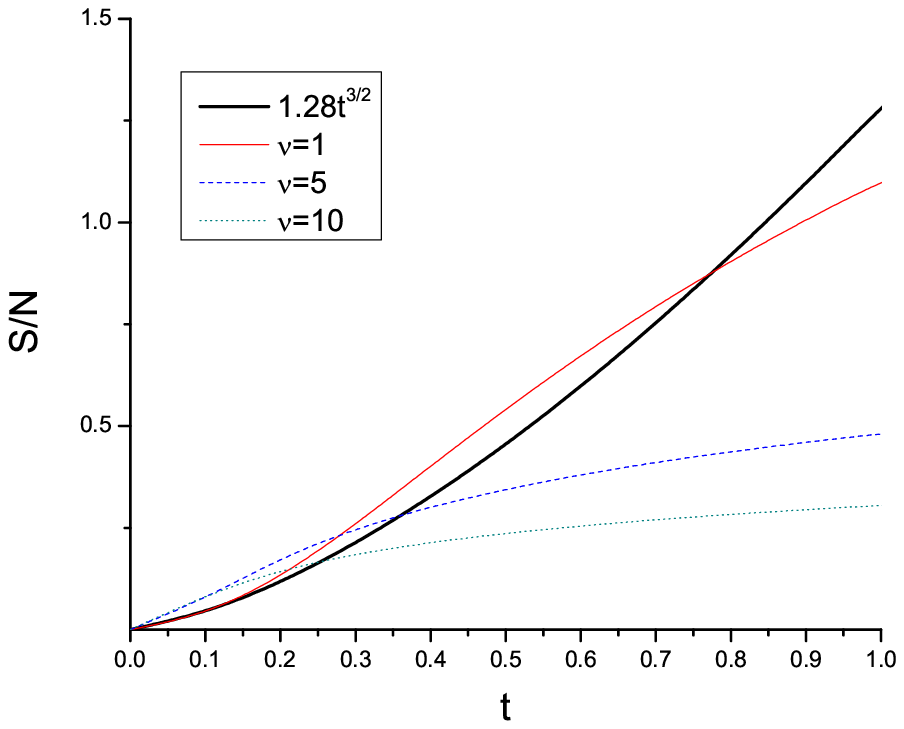}
\end{center}
\caption{(Color online)
The same as in Fig.3 but for the entropy per particle, $S/N$. It is seen that
$t^{3/2} $ power law does not hold even at $\nu=1$.
 }
\label{fig5}
\end{figure}

In Fig.5 the entropy per particle, $[S/N](t,\nu)$, of the condensed
 noninteracting atoms in  optical lattice
is presented for various $\nu$. One can again see that the scaling relation
 $[S/N](t) \sim t^{3/2}$ which holds for IHB gas does not hold for the system of noninteracting
 atoms in optical lattices. For small filling factors $[S/N](t,\nu)$ is almost
 linear in the temperature. As for the $\nu$ dependence of this function, the entropy
 decreases as $1/\nu$ with increasing $\nu$. A similar behavior of $[S/N](t,\nu)$
 was found by Blakie and Porto \ci{pra69} where the authors studied entropy -
 temperature curves in a large scale of temperature for $\nu \leq 4$.

 There is a simple relation between the internal  energy and the pressure $P$:
 \be
 E=\dsfrac{d}{2}PV
 \lab{pv}
 \ee
which holds for IHB gas. In textbooks it is usually derived by the integration by parts
 of the free energy, $\Omega=-PV$. On the other hand, it can be
  shown that  \ci{mancarella}, this relation is a consequence
of the scale invariance of the Hamiltonian with respect to the dilation of coordinates, i.e.,
$r\rightarrow \lambda r$. It is clear that the presence of the periodic potential \re{1.1}
 breaks this invariance and hence the relation \re{pv} is not valid in  optical lattices.

   \section{ Conclusion}

    We have studied the critical temperature for ideal Bose gases loaded into the cubic
   periodic lattice potential without a harmonic trap. We obtained the transition temperatures,
   $T_{c}^{0}(\nu)/J$,  with arbitrary filling factors
    including the high filling limit. In one dimension  this limit corresponds to the quantum rotor regime.
    We have proposed a simple interpolating formula for $T_{c}^{0}(\nu)$  that  approximates
    the transition temperatures for one, two and three  spatial
     dimensions. It was  shown that $T_{c}^{0}/J$ is linear
    for $\nu \gtrsim 5$ at  any realistic dimensions. In contrast to the ideal homogenious Bose
    gases,
    the ones in optical lattices exibit a linear temperature dependence of the condensed fraction and the energy
    per particle, especially at large filling factors. We have also
    shown that  thermodynamic quantities such as the energy, heat capacity and the entropy , of the system of even noninteracting
     atoms in optical lattices depend not only on the reduced temperature,
     but also on the density, i.e., on filling factor $\nu=\rho a^d$.

      These results may  serve as a checkpoint
    for various experiments on optical lattices as well as theoretical studies of weakly interacting bose
    systems in periodic potentials.
    %The study ideal atomic Bose gases in the normal phase
    %is on progress.
\section*{Acknowledgments}
A.R. is indebted to the Abdus Salam International Center for Theoretical Physics (ICTP,
Laboratory Condensed Matter Physics ) for the hospitality
    during his visit and to E.Ya. Sherman  for useful discussions.
    This work is supported by Scientific and Technological Research Council of Turkey (T\"{U}BITAK),  under Grant B\.{I}DEB -2221.

%\newpage

\bb{99}
%%%%%%%%%%%%
\bi{lewenstein} M. Lewenstein, A. Sanpera, and V. Ahufinger, {\it
Ultracold atoms in optical lattices: Simulating quantum many-body
systems} (Oxford University Press, Oxford, 2012).
\bi{zwerger}W. Zwerger, J. Opt. B {\bf 5},S9 (2003)
\bi{stoofbook} H.T.C. Stoof, K.B. Gubbels, and D.B.M. Dickerscheid,
{\it Ultracold Quantum Fields} (Springer, Berlin, 2009).
\bi{Yukalovobsor} V. I. Yukalov, Laser Physics  {\bf 19}, 1 (2009);\\
V. I. Yukalov, Condensed Matter Physics, 16, 23002
(2013).
\bi{bailie} D. Baillie and  P. B. Blakie, Phys. Rev. {\bf A} 80,
031603(R) (2009)
\bi{mahmud}K. W. Mahmud, E. N. Duchon, Y. Kato, N. Kawashima, R. T. Scalettar, N. Trivedi, Phys. Rev.
 {\bf B} 84, 054302 (2011)
\bi{svistun}B. Capogrosso-Sansone, N. V. Prokofev, and B. V.
Svistunov, Phys. Rev. {\bf B} 75, 134302 (2007).
 \bi{trotzky}
S. Trotzky, L. Pollet, F. Gerbier, U. Schnorrberger, I. Bloch, N. V.
Prokofev, B. Svistunov and M. Troyer, Nature Phys. {\bf 6}, 998
(2010).
\bi{ourknr2}    H.Kleinert, Z. Narzikulov, A. Rakhimov,
J.  Stat.  Mech.: Theor.  Exp.  P01003 (2014).
[10] V. I. Yukalov, A. Rakhimov,
\bi{ouryukalov}   V. I. Yukalov, A.    Rakhimov   and S.  Mardonov,  Laser Phys. , 21,  264 (2011)
\bi{landau}L. D. Landau and E. M. Lifshitz, E. M. Statistical Physics, Part
1 (Pergamon Press, Oxford, 1980); E. M. Statistical Physics,
Part 2 (Pergamon Press, Oxford, 1980).
\bi{yuklpl}  V. I. Yukalov, Laser Phys. Lett. {\bf 1}, 435 (2004)
 \bi{dalfavo}  F.Dalfovo, S.Giorgini, L. Pitaevskii, and S.Stringari, Rev. Mod. Phys. 71, 463
 (1999)
  \bi{ourknr1} H. Kleinert, Z. Narzikulov, A. Rakhimov Phys. Rev. A {\bf 85},
063602 (2012).
 \bi{danshita}I. Danshita and P. Naidon , Phys. Rev. A
{\bf 79}, 043601  (2009).
\bi{pra69} P. B. Blakie and J. V. Porto, Phys. Rev. A {\bf 69} , 013603 (2004)
\bi{mancarella}
 F. Mancarella, G. Mussardo and A. Trombettoni, Nucl. Phys. B {\bf 887} (2014).
 \eb
\edc